\documentclass[11pt]{article}

\usepackage{amsmath}
\usepackage{graphicx}
\usepackage{indentfirst}
\usepackage{cite}

\begin{document}

\title{\bf{Cosmic Censorship Conjecture in Kerr-Sen Black Hole}}

\date{}
\maketitle

\begin{center}
\author{Bogeun Gwak}$^a$\footnote{rasenis@sejong.ac.kr}\\

\vskip 0.25in
$^{a}$\it{Department of Physics and Astronomy, Sejong University, Seoul 05006, Republic of Korea}\\
\end{center}
\vskip 0.6in

{\abstract
{The validity of the cosmic censorship conjecture for the Kerr-Sen black hole, which is a solution to the low-energy effective field theory for four-dimensional heterotic string theory, is investigated using charged particle absorption. When the black hole absorbs the particle, the charge on it changes owing to the conserved quantities of the particle. Changes in the black hole are constrained to the equation for the motion of the particle and are consistent with the laws of thermodynamics. Particle absorption increases the mass of the Kerr-Sen black hole to more than that of the absorbed charges such as angular momentum and electric charge; hence, the black hole cannot be overcharged. In the near-extremal black hole, we observe a violation of the cosmic censorship conjecture for the angular momentum in the first order of expansion and the electric charge in the second order. However, considering an adiabatic process carrying the conserved quantities as those of the black hole, we prove the stability of the black hole horizon. Thus, we resolve the violation. This is consistent with the third law of thermodynamics.
}
}

\thispagestyle{empty}
\newpage
\setcounter{page}{1}
\section{Introduction}
Black holes generate gravitational waves. Such waves are detected at the Laser Interferometer Gravitational-Wave Observatory (LIGO), where it has been shown that there are many massive black holes in our Universe\cite{Abbott:2016blz,Abbott:2016nmj}. Using gravitational waves, the properties of black holes can be observed and analyzed.

In various gravity theories, black holes are solutions to the field equations in four dimensions. In Einstein's gravity theory, the Schwarzschild black hole is a static solution, and the Kerr black hole is a solution to a rotating black hole. When coupled with the Maxwell field, a rotating black hole is represented by a Kerr-Newman black hole. In string theory, black holes can be coupled with other fields, such as the dilaton field, Yang-Mills field, and antisymmetric tensor gauge field. The Kerr-Sen black hole is obtained using the low-energy effective field theory for the heterotic string theory in four dimensions\cite{Sen:1992ua} and is characterized by conserved quantities such as mass, angular momentum, and electric charges, which are similar to those of the Kerr-Newman black hole. However, the geometry of the Kerr-Sen black hole is different from that of the Kerr-Newman black hole.

The mass of a black hole consists of reducible and irreducible energies. Reducible energy, such as the rotational energy of a Kerr black hole, can be extracted from a black hole via the Penrose process\cite{Bardeen:1970zz,Penrose:1971uk}. However, irreducible energy cannot be extracted by a physical process and increases even if the mass of the black hole decreases\cite{Christodoulou:1970wf,Christodoulou:1972kt}. The behavior of the irreducible energy is similar to that of the horizon area of the black hole, which is proportional to the Bekenstein-Hawking entropy\cite{Bekenstein:1973ur,Bekenstein:1974ax} because the irreducible energy is in the horizon area\cite{Smarr:1972kt}. The temperature of a black hole can be defined as a Hawking temperature because emissions radiate from the surface of the horizon owing to a quantum effect\cite{Hawking:1974sw,Hawking:1976de}. The laws of thermodynamics are constructed using the entropy and temperature. The laws of thermodynamics can be applied to gravitational radiation that is produced when black holes collide. After the collision of two Schwarzschild black holes, the entropy of the final system should be greater than that of the initial state. This is the second law of thermodynamics. The upper bounds of the radiation can then be obtained for the Schwarzschild black hole\cite{Hawking:1971tu}. 
Radiation energy is affected not only by the mass of black holes but also by the potential energy from the spin interaction between black holes. This has been studied in Kerr black holes\cite{Schiff:1960gi,Wilkins:1970wap,Mashhoon:1971nm,Wald:1972sz} and Kerr-(anti-)de Sitter black holes\cite{Gwak:2016icd}. In addition, the radiation may depend on the instability of the final black hole, which is related to the angular momentum of black holes\cite{Ahn:2014fwa,Kokkotas:2015uma,Gwak:2016cbq}. The precise values and waveforms of the gravitational wave from a collision can be analyzed using numerical relativity\cite{Smarr:1976qy,Smarr:1977fy,Sperhake:2015siy,Barkett:2015wia,Hinderer:2016eia}.

The solution for a black hole has a curvature singularity in the event horizon. The observation of the singularity is prevented by the cosmic censorship conjecture. There are two versions of the cosmic censorship conjecture--the weak and strong versions. Both versions are mathematically independent of each other. In this paper, we focus on the weak cosmic censorship conjecture (WCCC). The WCCC means that a visible singularity does not exist, so that the horizon of the black hole should stably cover its singularity\cite{Penrose:1969pc}. For Einstein's theory of gravity, the conjecture has been tested in a Kerr black hole by adding a particle\cite{Wald:1974ge}. Various other tests of the conjecture have also been conducted for Einstein gravity, including higher dimensions\cite{Dadhich:1997rq,Hubeny:1998ga,Jacobson:2009kt,Gwak:2011rp,Barausse:2011vx,Colleoni:2015afa,Hod:2016hqx,Natario:2016bay,Horowitz:2016ezu}. Conserved charges from black holes can change infinitesimally when a particle is absorbed into a black hole\cite{Gwak:2012hq,Gwak:2015sua}. These changes can also be used to test the conjecture. Constrained by the energy of the particle, the change should be consistent with the laws of thermodynamics, and the black hole cannot be overcharged by the particle absorption. Thus, the conjecture is valid in the Kerr-AdS black hole\cite{Gwak:2015fsa}. The cosmic censorship conjecture has not been well studied in black hole solutions for various gravity theories, such as the string theory. The conjecture of the Kerr-Sen black hole was violated using the numerical approach  in the second order of expansion to near-extremal condition\cite{Siahaan:2015ljs}.

In this paper, we will analytically evaluate the stability of the outer horizon using charged particle absorption in the Kerr-Sen black hole for the low-energy effective field theory of the heterotic string theory in four dimensions. The stability of the horizon is a necessary condition for the validity of WCCC, so that we cannot ensure WCCC without the stability of the horizon. In addition, WCCC was given in Einstein's gravity theory, and the validation of the black holes based on string theory would be an interesting finding. The equation of motion, which was obtained from the separability of the Hamilton-Jacobi method, was consistent with the laws of thermodynamics for the Kerr-Sen black hole. Changes in the black hole when it absorbs the charged particle are given in terms of momenta and electric charge of the particle. We have proven the stability of the horizon for a nonextremal black hole by showing the increase in the horizon area. Further, to test the stability for the extremal black hole, the changes in the black hole will be applied and it will be shown that the black hole cannot be overcharged by charged particle absorption. We also consider the near-extremal black hole to complete our analysis. In the near-extremal condition, the horizon is unstable via the charged particle absorption. The instability is found for the angular momentum of the particle in the first-order expansion of the near-extremal condition and for the electric charge in the second order. However, the instabilities are resolved in the consideration of the adiabatic process carrying the conserved quantities of the particle to the black hole. Hence, the black hole can be more near-extremal than before the charged particle absorption, but it cannot be overcharged. Thus, the horizon of the Kerr-Sen black hole is stable for the charged particle absorption.

This paper is organized as follows. In Sec.~\ref{sec2}, the Kerr-Sen black hole and its thermodynamic properties are briefly reviewed. In Sec.~\ref{sec3}, using the equation of motion obtained using the Hamilton-Jacobi method, the change in the Kerr-Sen black hole is written in terms of the conserved quantities of the charged particle and related to the laws of thermodynamics. In Sec.~\ref{sec4}, an explanation of the validity of the cosmic censorship conjecture based on the Kerr-Sen black hole absorbing a charged particle is provided. In Sec.~\ref{sec5}, we find the violation of the cosmic censorship conjecture in the near-extremal cases, but we resolve the violation using the adiabatic process. In Sec.~\ref{sec6}, the results are summarized.

\section{The Kerr-Sen Black Hole}\label{sec2}
The Kerr-Sen black hole is a solution to the low-energy effective field theory for heterotic string theory in four dimensions\cite{Sen:1992ua}. Heterotic string theories are ten-dimensional theories with $\mathcal{N}=1$ supersymmetry constructed by combining left-moving excitations of the 26-dimensional bosonic string theory with the right-moving excitations of the ten-dimensional superstring theory. There are two constructions of the heterotic string. One is a superstring with a {\it{E}$_8\times$\it{E}$_8$} gauge group. The other is with the {\it{SO}(32)} gauge group. The bosonic part of the effective action of the heterotic string theories is given in the four-dimensional string frame\cite{Sen:1992ua,Houri:2010fr},
\begin{align}\label{eq:action1}
S=-\int d^4 x \sqrt{-g}e^\Phi\left(-R+\frac{1}{12}H_{\mu\nu\rho}H^{\mu\nu\rho}-g^{\mu\nu}\partial_\mu\Phi\partial_\nu\Phi+\frac{1}{8}F_{\mu\nu}F^{\mu\nu}\right)\,,
\end{align}
where the metric is $g_{\mu\nu}$ in the string frame, and the dilaton field is given as $\Phi$. The action in Eq.~(\ref{eq:action1}) contains\cite{Siahaan:2015ljs}
\begin{align}
H_{\mu\nu\rho}&=\partial_\mu B_{\nu\rho}+\partial_\rho B_{\mu\nu}+\partial_\nu B_{\rho\mu}-\frac{1}{4}\left(A_\mu F_{\nu\rho}+A_\rho F_{\mu\nu}+A_\nu F_{\rho\mu}\right)\,,\quad F_{\mu\nu}=\partial_\mu A_{\nu}-\partial_\nu A_{\mu}\,,
\end{align}   
where the Maxwell field and antisymmetric tensor gauge field are $A_{\mu}$ and $B_{\mu\nu}$. In the heterotic string theory, charged black hole solutions can be generated from neutral solutions by applying the Hassan-Sen transformation\cite{Hassan:1991mq}. The Kerr-Sen black hole is also a solution generated from the Kerr black hole; therefore, the field components are different from the Einstein-Maxwell theory providing the Reissner-Nordstr{\"{o}}m black hole. The solution for the Kerr-Sen black hole carrying conserved quantities, such as mass $M$, angular momentum $J$, and electric charges $Q$, is obtained in the Einstein frame\cite{Sen:1992ua}.
\begin{align}\label{eq:metricKdS}
ds^2&=-\frac{\Delta_r}{\rho_b^2}\left(dt-a\sin^2\theta d\phi\right)^2+\frac{\rho_b^2}{\Delta_r}dr^2+\rho_b^2 d\theta^2+\frac{\sin^2\theta}{\rho_b^2}\left(a\,dt-\left(r^2+a^2+2br\right)d\phi\right)^2\,,\\
\rho^2&=r^2+a^2\cos^2\theta\,,\quad\rho_b^2=\rho^2+2br\,,\quad\Delta_r=r^2+a^2-2(M-b)r\,,\quad b=\frac{Q^2}{2M}\,,\quad J=Ma\,,\nonumber
\end{align}
where the spin parameter is written as $a$, and the Maxwell field components are
\begin{eqnarray}
 A_t=-\frac{Qr}{\rho_b^2}\,,\quad A_\phi=\frac{aQr\sin^2\theta}{\rho_b^2}\,.
\end{eqnarray}
The inner and outer horizons are written as
\begin{align}\label{eq:horizon3}
r_{in}=M-\frac{Q^2}{2M}-\sqrt{\left(M-\frac{Q^2}{2M}\right)^2-\left(\frac{J}{M}\right)^2}\,,\quad r_{h}=M-\frac{Q^2}{2M}+\sqrt{\left(M-\frac{Q^2}{2M}\right)^2-\left(\frac{J}{M}\right)^2}\,.
\end{align}
The spin parameter is bounded as
\begin{align}\label{eq:bounded1}
|a| \leq \left|M-\frac{Q^2}{2M}\right|\,.
\end{align}
The equality in Eq.~(\ref{eq:bounded1}) gives maximally saturated charges as an extremal condition. For the extremal black hole, the inner and outer horizons correspond to
\begin{align}
r_e=M-\frac{Q^2}{2M}\,.
\end{align}
The horizon area $A_H$ of the extremal Kerr-Sen black hole is interestingly given as
\begin{align}
A_H=4\pi (r_e^2+a^2+2br_e)=8\pi |J|\,,
\end{align}
where the area only depends on the angular momentum. Further, this form of the area for the extremal condition is identical to that of a different rotating charged black hole in the five-dimensional framework\cite{Sen:1992ua,Frolov:1987rj}. The angular velocity at the outer horizon is
\begin{align}
\Omega_h=\frac{a}{r_h^2+a^2+2br_h}\,.
\end{align} 
The Hawking temperature, Bekenstein-Hawking entropy, and electric potential are given as
\begin{align}
T_h=\frac{Q^2-2M(M-r_h)}{8\pi M^2 r_h}\,,\quad S_{BH}=\frac{1}{4}A_H=\pi (r_h^2+a^2+2br_h)\,,\quad \Phi_h=\frac{Q}{2M}\,.
\end{align}
Then, the first law of thermodynamics is described as
\begin{align}
dM=T_h dS_{BH}+\Omega_h dJ +\Phi_h dQ\,.
\end{align}
There is a point to be noted in the work done in \cite{Siahaan:2015ljs} which investigated WCCC in the Kerr-Sen black hole. In \cite{Siahaan:2015ljs}, the violation of WCCC is numerically observed in near-extremal black holes for the second-order expansion. However, we analytically found the violation in both the first and second orders in this work. Further, in the consideration of the adiabatic process, we resolve the violations. This is consistent with the third law of thermodynamics. Finally, we have concluded that the Kerr-Sen black hole cannot be overspun by a particle.

\section{Thermodynamics via Charged Particle Absorption}\label{sec3}
Owing to the conserved quantities of an absorbed particle, the Kerr-Sen black hole undergoes changes that should be constrained by the equation of particle motion. To describe these changes in terms of the conserved quantities of the particle, the first-order equation of motion is obtained using the Hamilton-Jacobi method. The charged particle, with energy $E$ and angular momentum $L$, from the black hole is considered, and the Hamilton-Jacobi action of the particle is written as
\begin{align}
\quad S=\frac{1}{2}m^2\lambda-Et+L\phi+S_r(r)+S_\theta(\theta)\,,
\end{align}
where the mass of the particle is $m$, and the affine parameter is $\lambda$. The Hamiltonian of the particle with momentum $p_\mu$ and electric charge $e$ is given as
\begin{align}
\mathcal{H}=\frac{1}{2}g^{\mu\nu} (p_\mu-eA_\mu)( p_\nu-eA_\nu)\,.
\end{align}
The Hamiltonian equation is separable\cite{Hioki:2008zw,Houri:2010fr}, and all of the equations of motion can be obtained. However, the radial and $\theta$-directional equations of motion are necessary for the case of charged particle absorption. The equations of motion in terms of the radial and $\theta$-directional momenta, $p^r$ and $p^\theta$ are
\begin{align}\label{eq:eqmt01}
p^r&=\frac{d r}{d \lambda}=\frac{\Delta_r}{\rho_b^2}\sqrt{R(r)}\,,\quad p^\theta=\frac{d \theta}{d \lambda}=\frac{1}{\rho_b^2}\sqrt{\Theta(\theta)}\,,\\
R(r)&=-\frac{1}{\Delta_r}\left(\mathcal{K}+m^2r^2+2bm^2r-\frac{1}{\Delta_r}\left[(r^2+a^2+2br)(-E)+aL+eQr\right]^2\right)\,,\nonumber\\
\Theta(\theta)&=\mathcal{K}-a^2m^2\cos^2\theta-\sin^2\theta\left[-aE+L\csc^2\theta\right]^2\,,\nonumber
\end{align}
where the separate variable is $\mathcal{K}$. The particle is assumed to be absorbed into the black hole when it passes through the outer horizon. Thus, the relationship between conserved quantities of the particle at the horizon must be determined. For a given location, the energy of the particle can be written in terms of momenta and electric charge by removing the separate variable $\mathcal{K}$ in Eq.~(\ref{eq:eqmt01}). The energy equation is given as
\begin{align}\label{eq:energyeq01}
\alpha E^2 +2\beta E +\gamma=0\,,
\end{align}
where
\begin{align}
&\alpha = (a^2+r(2b+r))^2-a^2\Delta_r \sin^2\theta\,,\quad \beta=aL\Delta_r-(aL+eQr)(a^2+r(2b+r))\,,\nonumber\\
&\gamma=a^2L^2+eQr(eQr+2aL)-m^2 r (2b+r)\Delta_r -((p^r)^2+(p^\theta)^2\Delta_r)\rho_b^4-\Delta_r(a^2 m^2 \cos^2\theta+L^2\csc^2\theta)\,.\nonumber
\end{align}
A positive root is chosen in Eq.~(\ref{eq:energyeq01}) because it describes a future-forwarding particle\cite{Christodoulou:1970wf,Christodoulou:1972kt}. Since the energy is the conserved quantity for the time translation symmetry, the positive sign represents a particle forwarding to the black hole in a positive time flow. By the way, the root of the negative sign indicates a particle in a past-forward time, so the case represents the particle outgoing from the black hole. The root represents the relationship between the conserved quantities of the particle for a given location. The particle is assumed to be absorbed into the black hole when it passes through the outer horizon, where the energy of the particle $E_h$ can be written in terms of momenta and electric charge as
\begin{align}\label{eq:particlecon}
E_h=\frac{aL+eQr_h}{r_h(2b+r_h)+a^2}+\frac{\rho_b^2|p^r|}{r_h(2b+r_h)+a^2}\,,
\end{align}
where there is no $\theta$-directional dependency, and the energy of the particle is written in terms of the conserved quantities and radial momentum of the particle. The energy of the particle is not a variable now, but a value of a function determined by the momenta and electric charge of the particle. These conserved quantities change those of the black hole, so that
\begin{align}\label{eq:changecon}
\delta M =E_h\,,\quad \delta J = L\,,\quad \delta Q = e\,.
\end{align}
Changes in the black hole are related to Eq.~(\ref{eq:particlecon}) imposed by Eq.~(\ref{eq:changecon}). Thus, the relationship between the changes is given as
\begin{align}\label{eq:constrained}
\delta M = \frac{a\delta J +Qr_h\delta Q}{r_h(2b+r_h)+a^2}+\frac{\rho_b^2|p^r|}{r_h(2b+r_h)+a^2}\,.
\end{align}
Therefore, the changes should be constrained by Eq.~(\ref{eq:constrained}) during charged particle absorption. The changes in $\delta M$, $\delta J$, and $\delta Q$ of the black hole caused by the absorption will be related to each other by Eq.~(\ref{eq:constrained}). In addition, the radial momentum is also related to another thermal property. To test the stability of the horizon, first the variation of the horizon $r_h$ can be shown under Eqs.~(\ref{eq:metricKdS}) and (\ref{eq:constrained}), and then
\begin{align}\label{eq:deltarh1}
\delta \Delta_r&=\delta\left(r_h^2+\left(\frac{J}{M}\right)^2-2(M-\frac{Q^2}{2M})r_h\right)=0\,,\\
\delta r_h &=\frac{J(2J^2+M(Q^2-2M^2)r_h)}{2M^4 r_h (Q^2+2M(r_h-M))}L+\frac{Q(2J^2+M(Q^2-2M^2)r_h)}{2M^3 (Q^2+2M(r_h-M))}e+\frac{(2J^2+M(2M^2+Q^2)r_h)}{2M^3 r_h (Q^2+2M(r_h-M))}|p^r|\,,\nonumber
\end{align}
where there is freedom to choose the sign of $L$ and $e$; hence, it is still difficult to determine the change in $r_h$. To validate WCCC, the surface of the horizon should cover the inside of the black hole. Therefore, we just illustrate how the surface area of the black hole changes under particle absorption. As shown in Eq.~(\ref{eq:constrained}), the angular momentum and electric charge can make a negative contribution to the mass of the black hole; hence, one can expect the surface area to shrink to the extremal bound of the black hole, and then, it can disappear. To test the change in the area, we investigated the change in the Bekenstein-Hawking entropy, which is proportional to the area of the horizon. The change can be obtained from
\begin{align}\label{eq:changeent01}
&\delta S_{BH}=\delta \left(\pi (r_h^2+a^2+2br_h)\right)=\frac{\partial S_{BH}}{\partial J}\delta J+\frac{\partial S_{BH}}{\partial Q}\delta Q+\frac{\partial S_{BH}}{\partial M}\delta M+\frac{\partial S_{BH}}{\partial r_h}\delta r_h\,,
\end{align}
where
\begin{align}
\frac{\partial S_{BH}}{\partial J}=\frac{2\pi J}{M^2}\,,\quad \frac{\partial S_{BH}}{\partial Q}=\frac{2\pi r_h Q}{M}\,,\quad \frac{\partial S_{BH}}{\partial M}=-\frac{\pi(2J^2+MQ^2r_h)}{M^3}\,,\quad \frac{\partial S_{BH}}{\partial r_h}=\frac{\pi(Q^2+M r_h)}{M}\,.
\end{align}
Inserting Eqs.~(\ref{eq:constrained}) and (\ref{eq:deltarh1}) to replace $\delta M$ and $\delta r_h$ in terms of $\delta J$ and $\delta Q$, we find the increase in entropy as
\begin{align}\label{eq:changeent}
&\delta S_{BH}=\frac{4\pi M\rho_b^2 |p^r|}{Q^2-2M(M-r_h)}>0\,.
\end{align}
The change in entropy is proportional to the radial momentum, which is always positive for the incoming particle. Therefore, since there is no limitation on choosing the sign of $L$ and $e$, the surface area always increases, whatever particle is absorbed by the black hole. In addition, it also means that the entropy of the black hole always increases so that the second law of thermodynamics is satisfied during charged particle absorption. From Eq.~(\ref{eq:changeent}), we can find out that there is a close relationship between the stability of the horizon and the second law of thermodynamics. The second law of thermodynamics is related to the behavior of the irreducible mass of the black hole, which is an increasing part of the mass. Since the mass of the black hole consists of the irreducible mass and reducible energy, even if the total mass of the black hole decreases due to the decrease in reducible mass, the irreducible mass should increase during particle absorption. We can expect the change in the mass in Eq.~(\ref{eq:constrained}) to explicitly include reducible energies related to $\delta J$ and $\delta e$, and the remaining terms may be related to the irreducible mass. After removing the $\delta J$ and $\delta e$ parts in $\delta M$ in Eq.~(\ref{eq:constrained})\cite{Christodoulou:1970wf,Christodoulou:1972kt}, the integration of the left-hand side gives the irreducible mass and its change,
\begin{align}
M_{ir}=\sqrt{r_h^2+a^2+2br_h}\,,\quad\delta M_{ir}&=\frac{2M\rho_b^2 |p^r|}{M_{ir}(Q^2-2M(M-r_h))}>0\,,
\end{align}
where the irreducible mass always increases during absorption. Since the radial momentum of the particle is proportional to the change in the irreducible mass, we can now describe the components of the black hole mass,
\begin{eqnarray}
M=M\left(M_{ir},\,M_{rot},\,M_{elec}\right)\,,
\end{eqnarray}
where the rotation energy $M_{rot}$ and electric energy $M_{elec}$ are reducible masses. In the particle absorption, each charge of the particle can perturb the corresponding charge of the black hole. Hence, when we investigate the stability of the horizon, all particle charges should be taken into consideration with the particle energy equation of Eq.~(\ref{eq:constrained}). Without these parts, the mass of the black hole can be easily increased or decreased, and consequently, overcharging of the black hole can be achieved. The particle energy in Eq.~(\ref{eq:constrained}) is not a specific relation only applied to this process, but a generally satisfied relation in black hole physics. The radial momentum can be rewritten as entropy using Eq.~(\ref{eq:changeent}) so that
\begin{align}
\delta M = T_h \delta S_{BH}+\Omega_h \delta J+\Phi_h \delta Q\,,
\end{align}
which is the first law of thermodynamics for the Kerr-Sen black hole. Changes in the black hole satisfy the laws of thermodynamics under charged particle absorption. Therefore, the outer horizon is stable in Kerr-Sen black hole as shown in the increase of the surface area for any particle satisfying the equation of motion. Further, we will show the stability of the horizon in the extremal Kerr-Sen black hole.

\section{Cosmic Censorship Conjecture in Extremal Kerr-Sen Black Hole}\label{sec4}
Charged particle absorption can change the geometry of the black hole infinitesimally, owing to changes in the function $\Delta_r$ in the metric. Since the function $\Delta_r$ determines the extremality and locations of horizons, the horizon can disappear when the black hole is overcharged beyond the extremal condition. Spacetime then becomes a naked singularity. However, this is prevented by WCCC. Thus, the black hole should be tested to determine if overcharging or instability of the horizon is possible. To prove the stability of the horizon, the extremal Kerr-Sen black hole is investigated in this study. If the angular momentum or electric charge of the black hole slightly increases during particle absorption, the black hole may overcharge and become a naked singularity. This can be tested by changing the sign of the minimum of the function $\Delta_r$ during particle absorption. The function $\Delta_r$ of the extremal Kerr-Sen black hole has one minimum point at $r=r_e$, and the horizon is located at this minimum point. The extremal conditions for the function $\Delta_r$ are
\begin{align}
\Delta_e=&\Delta_r{\Big |}_{r=r_e}=0\,,\quad \partial_{r_e}\Delta_e=\frac{\partial \Delta_r}{\partial r}{\Big |}_{r=r_e}=0\,,\quad \partial^2_{r_e} \Delta_e=\frac{\partial^2 \Delta_r}{\partial r^2}{\Big |}_{r=r_e}>0\,.
\end{align}
Particle absorption infinitesimally changes the charges in the black hole and the function $\Delta_r$ from $(M,J,Q)$ to $(M+\delta M,J+\delta J,Q+\delta Q)$, which slightly moves the minimum point to $r_e+\delta r_e$. The moved minimum point $r_e+\delta r_e$ satisfies 
\begin{align}
&\partial_{r_e}\Delta_e+\delta \partial_{r_e}\Delta_e=\frac{\partial \partial_{r_e}\Delta_e}{\partial M}\delta M+\frac{\partial \partial_{r_e}\Delta_e}{\partial J}\delta J+\frac{\partial \partial_{r_e}\Delta_e}{\partial Q}\delta Q+\frac{\partial \partial_{r_e}\Delta_e}{\partial r_e}\delta r_e=0\,,\\
&\partial_{r_e}\Delta_e=-2\left(M-\frac{Q^2}{2M}+2r_h\right)\,,\quad \frac{\partial \partial_{r_e}\Delta_e}{\partial M}=-2\left(1+\frac{Q^2}{2M^2}\right)\,,\quad \frac{\partial \partial_{r_e}\Delta_e}{\partial J}=0\,,\nonumber\\
&\frac{\partial \partial_{r_e}\Delta_e}{\partial Q}=\frac{2Q}{M}\,,\quad \frac{\partial \partial_{r_e}\Delta_e}{\partial r_e}=2\,.\nonumber
\end{align}
Hence, the solution of $\delta r_e$ is obtained as
\begin{align}
\delta r_e=\frac{2M^2+Q^2}{4M^3}L -\frac{JQ}{2M^3}e +\frac{\rho_b^2(2M^2+Q^2)}{4JM^2}|p^r|\,,
\end{align}
which indicates the movement of the minimum point by the charged particle. The point is still the minimum, because
\begin{align}
\frac{\partial^2 \Delta_r}{\partial r^2}{\Big |}_{r=r_e+\delta r_e}=\frac{\partial \partial_{r_e}\Delta_e}{\partial r_e}>0\,.
\end{align}
At the point $r_e+\delta r_e$, the minimum value of the function $\Delta_r$ is obtained as
\begin{align}
\Delta_r+\delta \Delta_e& = \frac{\partial \Delta_e}{\partial M} \delta M+\frac{\partial \Delta_e}{\partial J} \delta J+\frac{\partial \Delta_e}{\partial Q} \delta Q+\frac{\partial \Delta_e}{\partial r_e} \delta r_e=-\frac{2\rho_b^2}{M}|p^r|<0\,,
\end{align}
where
\begin{align}
\frac{\partial \Delta_e}{\partial M}&=-\frac{2J^2+2M^3r_e+MQ^2r_e}{M^3}\,,\quad \frac{\partial \Delta_e}{\partial J}=\frac{2J}{M^2}\,,\quad \frac{\partial \Delta_e}{\partial Q}=\frac{2Q r_e}{M}\,,\quad \frac{\partial \Delta_e}{\partial r_e}=-2M+\frac{Q^2}{M}+2r_e=0\,.
\end{align}
The minimum value is always negative; hence, the black hole cannot be overcharged. Therefore, the outer horizon is stable during particle absorption, and the extremal condition is not satisfied. The black hole now has two horizons, an inner one and an outer one, which are represented as
\begin{align}
r_e+\delta r_\pm=\frac{J}{M}+\frac{2M^2+Q^2}{4M^3}\delta J -\frac{JQ}{2M^3}\delta Q +\frac{\rho_b^2(2M^2+Q^2)}{4JM^2}|p^r| \pm \sqrt{\frac{2\rho_b^2}{M}|p^r|}\,,
\end{align}
where the positive horizon is the outer horizon and the negative horizon is the inner horizon. The change in mass, which is increased because of particle absorption, is greater than that caused by rotational and electric energies, so that extremality is broken. Therefore, the horizon is stable for the Kerr-Sen black hole.

\section{Cosmic Censorship Conjecture in Near-Extremal Kerr-Sen Black Hole}\label{sec5}
We now consider charged particle absorption in the near-extremal black hole. The near-extremal condition is introduced in Eq.~(\ref{eq:horizon3}) and (\ref{eq:bounded1}). Then,
\begin{eqnarray}\label{eq:nearcond1}
\left(M^2-\frac{Q^2}{2}\right)^2-J^2=D\,,
\end{eqnarray}
where $D$ is the free parameter having a very small value owing to the near-extremal condition, $D\ll 1$. When the conserved quantities of the black hole, $(M,J,Q)$, change to $(M+\Delta M,J+\Delta J,Q+\Delta Q)$ by the absorption, if the value of $D$ becomes negative, the horizon disappears, and then the cosmic censorship conjecture is violated. In this section, we use $\Delta$ instead of $\delta$, because we will consider up to the second order in the expansion of $\Delta M$, $\Delta J$, $\Delta Q$, and $p^r$. The value of $D$ changes to $D+\Delta D$ in $(M+\Delta M,J+\Delta J,Q+\Delta Q)$, and the first order of the expansion in Eq.~(\ref{eq:nearcond1}) is given as
\begin{align}
D+\Delta D &=\left((M+\Delta M)^2-\frac{(Q+\Delta Q)^2}{2}\right)^2-(J+\Delta J)^2 \\
&\approx D-2J\Delta J+4M\left(M^2-\frac{Q^2}{2}\right)\Delta M -2Q\left(M^2-\frac{Q^2}{2}\right)\Delta Q\,,
\end{align}
in which $\Delta M$ can be removed by Eq.~(\ref{eq:constrained}), because conserved quantities are carried by the charged particle. We can then remove $\Delta M$ in terms of the electric charge and momenta of the particle. This can be written in a compact form as
\begin{eqnarray}\label{eq:firstnearexpandsion1}
D+\Delta D =D-\frac{J\sqrt{D}}{M r_h}\Delta J+\frac{(2M^2-Q^2)\rho_b^2}{r_h}|p^r|\,,
\end{eqnarray}
in which there is no $\Delta Q$ term; hence, the angular momentum of the particle is dominant in the first-order expansion. When this becomes negative, the cosmic censorship conjecture is violated. For the simplest example, we set $p^r=0$. Then, if the particle has an angular momentum greater than
\begin{align}\label{eq:firstorder1}
L = \Delta J >\frac{\sqrt{D} M r_h}{J}\,,
\end{align} 
the value of Eq.~(\ref{eq:firstnearexpandsion1}) becomes negative, so the black hole is overcharged. Then, WCCC is violated. Note that we choose the value of $D$ to be small enough that the horizon of the near-extremal black hole can disappear with a small angular momentum of the particle satisfying Eq.~(\ref{eq:firstorder1}), so that the cosmic censorship conjecture is violated. However, Eq.~(\ref{eq:firstorder1}) implies that the scale of $\Delta J$ of the particle is of the order of $D$ of the black hole; hence, the scale of $\Delta J$ is large. We then need to consider the path from $(M,J)$ to $(M+\Delta M,J+\Delta J)$. The violation of the cosmic censorship conjecture is a result of a jump from $(M,J)$ to $(M+\Delta M,J+\Delta J)$ at a once. However, since the particle collides at a specific point of the horizon, the energy of the particle takes time to spread across the whole volume of the black hole. In the jump at a once, the energy is assumed to merge into the black hole at a moment. Hence, the jump at a once might be unphysical to represent this process. Instead of this, if we consider a continuous path called the adiabatic process\cite{Chirco:2010rq}, the result becomes different. We define $f(D,M,J)=\frac{J \sqrt{D} }{M r_h}$. Then, the continuous path is given in terms of $n$ steps
\begin{align}
&(D,M,J) \rightarrow (D_1,M+\delta M, J+\delta J) \rightarrow (D_2,M+2\delta M, J+2\delta J) \rightarrow...\\
&\rightarrow (D_n,M+n\delta M, J+n\delta J)\,,\nonumber
\end{align}
where $\delta J =\frac{\Delta J}{n}$ and $D_{i-1}>D_i$, because the black hole becomes more near-extremal than before a step. Hence, the final state of the black hole is also $(M+\Delta M, J+\Delta J)$. The near-extremal condition $D$ changes with each step.
\begin{align}\label{eq:continue1}
&D \rightarrow D_1=D- f(D,M,J)\delta J \rightarrow D_2=D_1-(D_1,M+\delta M, J+\delta J)\delta J \rightarrow...\\
&\rightarrow D_n=D_{n-1}-f(D_{n-1},M+(n-1)\delta M, J+(n-1)\delta J)\delta J= D-\sum_{i=0}^{n-1}f(D_i,M+i\delta M,J+i\delta J)\delta J\,,\nonumber
\end{align}
where $D_0=D$. Note that the case of $n=1$ represents the jump from $(M,J)$ to $(M+\Delta M,J+\Delta J)$. Since the horizon of the near-extremal condition is given as $r_h=\frac{J}{M}+\frac{\sqrt{D}}{M}$ from Eq.~(\ref{eq:horizon3}),
\begin{align}
f(D,M,J)=\frac{J\sqrt{D}}{J+\sqrt{D}}\sim\sqrt{D}\,,
\end{align}
which means that $f(D_{i-1},M+(i-1)\delta M,J+(i-1)\delta J)>f(D_i,M+i\delta M,J+i\delta J)$. Owing to the definition of $\delta J$,
\begin{align}\label{eq:continue3}
D_{n}=D_{n-1}-\frac{1}{n-1}f(D_{n-1},M+(n-1)\delta M, J+(n-1)\delta J)\Delta J\,.
\end{align}
If we consider a large enough number of steps $n \gg 1$ for a positive $D_{n-1}$, we always find $0<D_n<D_{n-1}$. Hence, the near-extremal condition $D_n$ can be positive. Therefore, the black hole cannot be overcharged by the particle absorption. This is consistent with the third law of thermodynamics. This is generalized to the second order of $\Delta Q$.

For the charged particle with the angular momentum, the second-order expansion of $\Delta Q$ is obtained from Eq.~(\ref{eq:nearcond1}) as
\begin{eqnarray}
D-\frac{(2M^2-Q^2)^2}{4M^2}\Delta Q^2+\frac{(2M^2 Q -Q^3)\rho_b^2}{2M^2 r_h}|p^r|\Delta Q+\frac{(6M^2-Q^2)\rho_b^4}{4M^2 r_h^2}|p^r|^2\,.
\end{eqnarray}
Therefore, in the second order of $\Delta Q$, the horizon of the near-extremal black hole can disappear. In this case, the particle having $p^r=0$ should have an electric charge $e=\Delta Q$ such that
\begin{eqnarray}
e>\frac{2M\sqrt{D}}{|2M^2-Q^2|}\,,
\end{eqnarray} 
in which the black hole can be overcharged by the charge of the particle. In a manner similar to Eqs.~(\ref{eq:continue1}) to (\ref{eq:continue3}), we define $n$ steps in the absorption, and the near-extremal condition $D_n$ of the $n$th step is obtained as
\begin{eqnarray}
&&D_{n}=D_{n-1}-\frac{1}{(n-1)^2}g(D_{n-1},M+(n-1)\delta M, J+(n-1)\delta Q)(\Delta Q)^2\,,\\
&&g(D,M,Q)=\frac{(2M^2-Q^2)^2}{4M^2}\,.\nonumber
\end{eqnarray} 
In the consideration of sufficiently large $n$, the value of $D_n$ is always positive. Therefore, the horizon is still stable in the second order of $\Delta Q$.

We consider an adiabatic process of the charged particle absorption to prove the stability of the horizon. This process can be generalized to arbitrary values of $L$ and $e$ for the particle. In these cases, we can also assume a large number of steps $n$, so that the near-extremal condition $D_n$ is not negative. Using this process, we can divide the conserved quantities of the particle making the horizon unstable to infinitesimally small steps, so the horizon is still stable in the Kerr-Sen black hole. This result is consistent with the third law of thermodynamics.

\section{Summary}\label{sec6}
In this study, the cosmic censorship conjecture of the Kerr-Sen black hole, which is a solution to the effective action of the low-energy string theory, was investigated. According to conjecture, the singularity of the black hole should be covered by its horizon, but the horizon may disappear when the black hole is overcharged beyond the extremal condition. Thus, the extremal black hole has been tested with maximally saturated charges for a given mass using charged particle absorption. The equation of motion for the charged particle was obtained using the Hamilton-Jacobi method and was separable to ensure that the energy of the particle, in terms of its momenta and electric charge at the outer horizon of the black hole, was exact. Further, the energy and charges of the particle were assumed to be absorbed into those of the black hole when the particle passed through the outer horizon of the black hole. The particle infinitesimally changes the angular momentum and electric charge of the black hole, and we can show the increase in the horizon area of the black hole. Hence, the black hole cannot be overcharged from a nonextremal black hole by the absorption. It also means that the Bekenstein-Hawking entropy increases, which obeys the second law of thermodynamics. The equation of motion can be rewritten as the first law of thermodynamics; hence, our approach is not a specific case, but a generally agreed-upon one. Therefore, changes in the black hole satisfy the laws of thermodynamics under charged particle absorption. Using the equation of motion, the stability of the horizon for the extremal Kerr-Sen black hole has been investigated, and during particle absorption, the extremal black hole becomes a nonextremal one with inner and outer horizons, because the mass of the black hole increases beyond that of its charges. In the near-extremal black hole, the instability of the horizon is found in the first- and second-order expansion of the near-extremal condition. However, if we consider a continuous path carrying the conserved quantities from the particle to the black hole, the instability is resolved, and the black hole still has the outer horizon. Thus, the horizon is stable in the Kerr-Sen black hole based on the effective action of the low-energy string theory.

\vspace{10pt} 

{\bf Acknowledgments}

This work was supported by the faculty research fund of Sejong University in 2016. BG was supported by Basic Science Research Program through the National Research Foundation of Korea(NRF) funded by the Ministry of Science, ICT \& Future Planning(NRF-2015R1C1A1A02037523).


\begin{thebibliography}{99}
\bibitem{Abbott:2016blz}
{\bf Virgo, LIGO Scientific} Collaboration, B.~P. Abbott et~al. {\em Phys. Rev.
  Lett.} {\bf 116}, 061102 (2016), no.~6.

\bibitem{Abbott:2016nmj}
{\bf Virgo, LIGO Scientific} Collaboration, B.~P. Abbott et~al. {\em Phys. Rev.
  Lett.} {\bf 116}, 241103 (2016), no.~24.

\bibitem{Sen:1992ua}
A.~Sen {\em Phys. Rev. Lett.} {\bf 69}, 1006--1009 (1992).

\bibitem{Bardeen:1970zz}
J.~M. Bardeen {\em Nature} {\bf 226}, 64--65 (1970).

\bibitem{Penrose:1971uk}
R.~Penrose and R.~M. Floyd {\em Nature} {\bf 229}, 177--179 (1971).

\bibitem{Christodoulou:1970wf}
D.~Christodoulou {\em Phys. Rev. Lett.} {\bf 25}, 1596--1597 (1970).

\bibitem{Christodoulou:1972kt}
D.~Christodoulou and R.~Ruffini {\em Phys. Rev.} {\bf D4}, 3552--3555 (1971).

\bibitem{Bekenstein:1973ur}
J.~D. Bekenstein {\em Phys. Rev.} {\bf D7}, 2333--2346 (1973).

\bibitem{Bekenstein:1974ax}
J.~D. Bekenstein {\em Phys. Rev.} {\bf D9}, 3292--3300 (1974).

\bibitem{Smarr:1972kt}
L.~Smarr {\em Phys. Rev. Lett.} {\bf 30}, 71--73 (1973). [Erratum: Phys. Rev.
  Lett.30,521(1973)].

\bibitem{Hawking:1974sw}
S.~W. Hawking {\em Commun. Math. Phys.} {\bf 43}, 199--220 (1975).

\bibitem{Hawking:1976de}
S.~W. Hawking {\em Phys. Rev.} {\bf D13}, 191--197 (1976).

\bibitem{Hawking:1971tu}
S.~W. Hawking {\em Phys. Rev. Lett.} {\bf 26}, 1344--1346 (1971).

\bibitem{Schiff:1960gi}
L.~I. Schiff {\em Proc. Nat. Acad. Sci.} {\bf 46}, 871 (1960).

\bibitem{Wilkins:1970wap}
D.~Wilkins {\em Annals of Physics} {\bf 61}, 277 -- 293 (1970), no.~2.

\bibitem{Mashhoon:1971nm}
B.~Mashhoon {\em J. Math. Phys.} {\bf 12}, 1075--1077 (1971).

\bibitem{Wald:1972sz}
R.~M. Wald {\em Phys. Rev.} {\bf D6}, 406--413 (1972).

\bibitem{Gwak:2016icd}
B.~Gwak and D.~Ro 
{{\tt
  arXiv:1610.04847}}.

\bibitem{Ahn:2014fwa}
W.-K. Ahn, B.~Gwak, B.-H. Lee, and W.~Lee {\em Eur. Phys. J.} {\bf C75}, 372
  (2015), no.~8.

\bibitem{Kokkotas:2015uma}
  K.~D.~Kokkotas, R.~A.~Konoplya and A.~Zhidenko {\em Phys. Rev. } {\bf D92},  064022 (2015), no.~6.

\bibitem{Gwak:2016cbq}
B.~Gwak and B.-H. Lee {\em JHEP} {\bf 07}, 079 (2016).

\bibitem{Smarr:1976qy}
L.~Smarr, A.~Cadez, B.~S. DeWitt, and K.~Eppley {\em Phys. Rev.} {\bf D14},
  2443--2452 (1976).

\bibitem{Smarr:1977fy}
L.~Smarr {\em Phys. Rev.} {\bf D15}, 2069--2077 (1977).

\bibitem{Sperhake:2015siy}
U.~Sperhake, E.~Berti, V.~Cardoso, and F.~Pretorius {\em Phys. Rev.} {\bf D93},
  044012 (2016), no.~4.

\bibitem{Barkett:2015wia}
K.~Barkett et~al. {\em Phys. Rev.} {\bf D93}, 044064 (2016), no.~4.

\bibitem{Hinderer:2016eia}
T.~Hinderer et~al. {\em Phys. Rev. Lett.} {\bf 116}, 181101 (2016), no.~18.

\bibitem{Penrose:1969pc}
R.~Penrose {\em Riv. Nuovo Cim.} {\bf 1}, 252--276 (1969). [Gen. Rel.
  Grav.34,1141(2002)].

\bibitem{Wald:1974ge}
R.~Wald {\em Annals Phys.} {\bf 82}, 548--556 (1974).

\bibitem{Dadhich:1997rq} 
N.~Dadhich and K.~Narayan {\em Phys. Lett. A} {\bf 231}, 335--338 (1997).

\bibitem{Hubeny:1998ga}
V.~E. Hubeny {\em Phys. Rev.} {\bf D59}, 064013 (1999).

\bibitem{Jacobson:2009kt}
T.~Jacobson and T.~P. Sotiriou {\em Phys. Rev. Lett.} {\bf 103}, 141101 (2009).
  [Erratum: Phys. Rev. Lett.103,209903(2009)].

\bibitem{Gwak:2011rp}
B.~Gwak and B.-H. Lee {\em Phys. Rev.} {\bf D84}, 084049 (2011).

\bibitem{Barausse:2011vx}
E.~Barausse, V.~Cardoso, and G.~Khanna {\em Phys. Rev.} {\bf D84}, 104006
  (2011).

\bibitem{Colleoni:2015afa}
M.~Colleoni and L.~Barack {\em Phys. Rev.} {\bf D91}, 104024 (2015).

\bibitem{Hod:2016hqx}
S.~Hod {\em Class. Quant. Grav.} {\bf 33}, 037001 (2016), no.~3.

\bibitem{Natario:2016bay}
J.~Natario, L.~Queimada and R.~Vicente {\em Class. Quant. Grav.} {\bf 33}, 175002 (2016), no.~17.

\bibitem{Horowitz:2016ezu}
G.~T. Horowitz, J.~E. Santos, and B.~Way {\em Class. Quant. Grav.} {\bf 33},
  195007 (2016), no.~19.

\bibitem{Gwak:2012hq}
B.~Gwak and B.-H. Lee {\em Class. Quant. Grav.} {\bf 29}, 175011 (2012).

\bibitem{Gwak:2015sua}
B.~Gwak and B.-H. Lee {\em Phys. Lett.} {\bf B755}, 324--327 (2016).

\bibitem{Gwak:2015fsa}
B.~Gwak and B.-H. Lee {\em JCAP} {\bf 1602}, 015 (2016).

\bibitem{Siahaan:2015ljs}
H.~M. Siahaan {\em Phys. Rev.} {\bf D93}, 064028 (2016), no.~6.

\bibitem{Houri:2010fr}
T.~Houri, D.~Kubiznak, C.~M. Warnick, and Y.~Yasui {\em JHEP} {\bf 07}, 055
  (2010).

\bibitem{Hassan:1991mq} 
  S.~F.~Hassan and A.~Sen,
  Nucl.\ Phys.\ B {\bf 375}, 103 (1992).

\bibitem{Frolov:1987rj} 
  V.~P.~Frolov, A.~I.~Zelnikov and U.~Bleyer,
  Annalen Phys.\  {\bf 44}, 371 (1987).

\bibitem{Hioki:2008zw}
K.~Hioki and U.~Miyamoto {\em Phys. Rev.} {\bf D78}, 044007 (2008).

\bibitem{Chirco:2010rq} 
  G.~Chirco, S.~Liberati and T.~P.~Sotiriou,
  Phys.\ Rev.\ D {\bf 82}, 104015 (2010).

\end{thebibliography}
\end{document}